\documentclass{article}
\usepackage{spconf,amsmath,graphicx,amsfonts}
\usepackage{multirow}
\usepackage{color}
\usepackage{enumitem}
\usepackage{booktabs}
\usepackage{multirow}
\usepackage{hhline}
\usepackage{amsmath}
\usepackage{algorithm}
\usepackage{algpseudocode}

\usepackage{hyperref}

\usepackage[T4,T1]{fontenc}
\makeatletter
\newcommand{\do@openE}[1]{%
  \mbox{\fontsize{#1}\z@\usefont{T4}{cmr}{m}{n}\symbol{130}}%
}
\newcommand{\openE}{\mathord{\mathchoice
  {\do@openE\tf@size}
  {\do@openE\tf@size}
  {\do@openE\sf@size}
  {\do@openE\ssf@size}
}}

\usepackage{ifthen}

\title{Robust DOA estimation from deep acoustic imaging}
\name{Adrian S. Roman$^{1}$\sthanks{corresponding author email: romanguz@usc.edu} \qquad Iran R. Roman$^{2}$ \qquad Juan P. Bello$^{2}$}
\address{$^{1}$ Viterbi School of Engineering, University of Southern California, California, USA \\
$^{2}$ Music and Audio Research Laboratory, New York University, New York, USA} 
\begin{document}
%\ninept
%
\maketitle
\begin{abstract}
Direction of arrival estimation (DoAE) aims at tracking a sound in azimuth and elevation. 
Recent advancements include data-driven models with inputs derived from ambisonics intensity vectors or correlations between channels in a microphone array.
A spherical intensity map (SIM), or acoustic image, is an alternative input representation that remains underexplored. 
SIMs benefit from high-resolution microphone arrays, yet most DoAE datasets use low-resolution ones.
Therefore, we first propose a super-resolution method to upsample low-resolution microphones. Next, we benchmark DoAE models that use SIMs as input. 
We arrive to a model that uses SIMs for DoAE estimation and outperforms a baseline and a state-of-the-art model. Our study highlights the relevance of acoustic imaging for DoAE tasks. 
\end{abstract}
\begin{keywords}
spherical intensity maps, acoustic imaging, sound event localization, super-resolution.
\end{keywords}
\section{Introduction}
\label{sec:intro}
% Direction of arrival estimation (DoAE) is the task to determine the location of sound sources in an environment by processing a signal by captured by a microphone \cite{adavanne2018sound, grumiaux2022survey}. %adavanne2019localization, guirguis2021seld, kapka2019sound}.
% % The recent release of large-scale DoAE datasets \cite{adavanne2019multi, politis2022starss22,lollmann2018locata} has allowed for an accelerated development of models \cite{adavanne2019localization,wang2023four,Liu_CQUPT_task3a_report}. 
% DoAE is an important component of assistive technologies for both low vision and audition individuals \cite{pandya2021ambient, georgiou2022someone}, and scene understanding in low-visibility conditions such as emergency response \cite{kim2020occupant,sun2021emergency}.

The release of large-scale sound event localization and detection (SELD) datasets \cite{shimada2023starss23,lollmann2018locata,politis_archontis_2022_6406873} allowed for deep learning approaches for direction of arrival estimation (DoAE) in favor of traditional %\cite{adavanne2019localization,wang2023four,Liu_CQUPT_task3a_report} 
%to evolve from 
beamforming \cite{grumiaux2022survey}.
Today, the best-performing models use a combination of two inputs: 1) the intensity vectors of first order ambisonics (FOA) and 2) the correlations between channels in a tetrahedral microphone (4 channels) \cite{wang2023four,hu2022}.
%Most models, including convolutional recurrent neural networks (CRNN) \cite{adavanne2018sound} and transformers \cite{wang2023four,Liu_CQUPT_task3a_report}, assume a 2D Euclidean azimuth and elevation, resulting in non-uniform localization errors \cite{kushwaha2022analyzing}. 
%Furthermore, they output vectors pointing to the location of sound events, and fail to capture features like loudness or resonance. 
%These factors result in models that are hard to interpret.% with underlying operations that are not interpretable. 
However, a spherical intensity map (SIM) is an alternative input representation that remains unexplored.

\begin{figure}

\includegraphics[width=8.5cm]{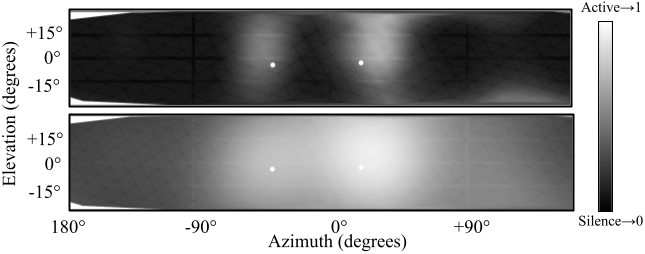}
\caption{The acoustic image by DeepWave (top) and DASB (bottom) for two sound sources. Dots denote ground truth.}
\label{fig:dw}
\end{figure}

% Perhaps the most interpretable representation is a %of sound sources is obtained by mapping them to a 
% spherical intensity field \cite{rafaely2015fundamentals}. 
A SIM is computed using delay-and-sum beamforming (DASB) \cite{rafaely2015fundamentals}. 
DeepWave, an {\it acoustic imaging} model \cite{simeoni2019deepwave}, builds upon DASB by applying sparse coding \cite{gregor2010learning} to deblur a cleaner representation of spatial sound activity (see Figure \ref{fig:dw}). % over the correlations between channels in a microphone array (Fig. \ref{fig:dw}). % shows example of DeepWave's highly-interpretable output.
% Graph Neural Networks (GNNs) offer interpretable operations remain largely under-explored for DoAE.
% DeepWave is a signal processing model with GNN operations, trained to carry out {\it acoustic imaging} \ref{}.
% Moreover, due to its use of graphical convolutions it can processing arbitrary microphone arrays with the high efficiency of its lightweight operations \cite{}. 
We evaluated whether we could use DeepWave's SIM to localize sound on the entire LOCATA dataset \cite{lollmann2018locata}, and we found that it is possible to achieve great localization performance (see Table \ref{tab:loc_results}, top row). 
However, DeepWave assumes high-resolution inputs, such as those captured by the EigenMike (32 channels) \cite{acoustics2013em32} or Pyramic (48 channels) \cite{scheibler2018pyramic} arrays. 

In this study we investigate whether high-resolution SIMs can be used to carry out DoAE. 
Since most SELD datasets use low-resolution arrays, we first demonstrate it is possible to upsample them using a super-resolution model.
%, upsampled to 32ch using a super-resolution model (see methods section), resulting in the same localization recall (LR) but suffering a hit in localization performance (see Table \ref{tab:loc_results}, 5th row).
% \begin{table}
% \begin{center}
% \begin{tabular}{llllll}
% \toprule\toprule
% Model & Input & $LE$ $\downarrow$ & $LR$ $\uparrow$ \\ 
% \midrule\midrule
%     \multirow{ 2}{*}{DW$\rightarrow \text{K-means}$} & em32ch & \textbf{14.8} & \textbf{99.2} \\ 
%     & tetra4ch $\rightarrow$ up32ch & 27.10  &  \textbf{99.2} \\ \hline
%    \bottomrule
%         \end{tabular}
%     \setlength{\belowcaptionskip}{-15pt}
%     \caption{Localization error (LE) and recall (LR) by DeepWave on the entire LOCATA dataset. Its acoustic imaging output is post-processed with K-means clustering to detect up to three simultaneous sound events (see methods). We evaluated two separate DeepWave models: one with a 32-channel input, and one with a 32-channel input upsampled from a four channel tetra microphone (see methods section). Legend: DW=DeepWave, up32ch=super-resolution upsampling.}
%     \label{tab:dw_results}
% \end{center}
% \end{table}
% In this study we investigate how low-resolution inputs impact DeepWave. 
Next, we benchmark models that use SIMs as input to carry out DoAE, and compare against a baseline and open-source state-of-the-art (SoTA) model.
Our code and data are openly-available.\footnote{https://github.com/adrianSRoman/DeepWaveDOA}
% is competitive for DoAE versus a baseline and the on the STARSS23 dataset \cite{shimada2023starss23}. 
In summary, our contributions are:
\begin{enumerate}[nolistsep]
% \item A dataset of spatial soundscapes with annotated sound event locations, in both em32ch and tetra4ch formats, generated using the em32ch room impulse responses METU-SPARG \cite{orhun_olgun_2019_2635758} and ARNI \cite{mckenzie2021dataset}.
\item A method to upsample the covariance between channels in a low-resolution microphone to high-resolution.% from low-resolution to high-resolution.% upsampling method using Complex-valued Deep Back Projection Networks (CDBPNs).
% \item A PyTorch implementation of the DeepWave model\footnote{https://github.com/adrianSRoman/DeepWaveTorch}.
% \item An ablation study to identify the optimal chain of DeepWave operations that yield the best DoAE performance. 
\item A benchmark of models that use SIMs for DoAE, evaluated on the LOCATA \cite{lollmann2018locata} and STARSS23 \cite{shimada2023starss23} datasets.
\end{enumerate}

\section{Related work}

Current DoAE models process the intensity vectors of FOA and correlations between channels in a tetrahedral microphone \cite{wang2023four,hu2022}.
% They trained on large datasets with moving sound events in real \cite{lollmann2018locata,shimada2023starss23} and simulated \cite{politis_archontis_2022_6406873} rooms. 
Significant ones include the ``Sound event localization and detection network'' (SELDnet) \cite{adavanne2018sound}, considered to be a baseline for the task, as well as SoTA models that use multi-head self-attention \cite{wang2023four} and multi-task outputs \cite{hu2022}.

A SIM is an acoustic image indicating the location of a sound source  \cite{chu2014robust}.
SIMs have not been assessed quantitatively as input for DoAE deep-learning models. 
% Acoustic imaging generates a visual representation .
The DASB algorithm \cite{rafaely2015fundamentals} is the most commonly-used method to compute SIMs, but resulting images exhibit poor angular resolution \cite{simeoni2019deepwave}.
DeepWave achieves an equivalent SIM using a \textit{backprojection} operation, and further denoises it using \textit{deblurring},
\begin{equation}
x^{(\ell)} = \tanh\Bigl(\underbrace{[\bar{\mathbf{B}} \circ \mathbf{B}]^{H} \text{vec}(\mathbf{\Sigma})}_{backprojection} + \underbrace{\mathbf{P}_{\theta} (\mathbf{L}) x^{(\ell - 1)}}_{deblurring} - \tau \Bigl),
\end{equation}
where $\mathbf{\Sigma} \in \mathbb{C}^{M\times M}$ is the instantaneous covariance matrix  of a microphone array with $M$ channels. 
\textit{Backprojection} is equivalent to DASB (sec. 5.2 in \cite{van2013signal}). 
It maps $\mathbf{\Sigma}\in\mathbb{C}^{M\times M}$ onto a uniformly-tiled SIM (i.e. tesselated \cite{kushwaha2022analyzing}), representing azimuth and elevation.
$\text{vec}(\mathbf{\Sigma})\in\mathbb{C}^{M^2}$ is a column-wise matrix-flattening operator, $\mathbf{B} \in \mathbb{C}^{M\times N}$ is a trainable matrix, and $\bar{\mathbf{B}}$ is its complex conjugate. 
$\circ$ is the Khatri-Rao product % is denoted by $\circ$, 
and $H$ denotes a Hermitian matrix.
\textit{Deblurring} iterates over an initial random spherical map $x^{(0)} \in \mathbb{R}^N$ using graphical convolutions that clean it.
$\mathbf{P}_{\theta} (\mathbf{L})$ is a polynomial of the graph Laplacian $\mathbf{L} \in \mathbf{R}^{NxN}$, and $\theta$ are graph convolution filters.
% proposes an alternative to DAS by applying sparse coding \cite{olshausen1996emergence, gregor2010learning} over the , where $M$ is the number of channels.

DeepWave applies these operations $L$ times with tanh for sparsity.
% , each with a bias $\tau$ and %the model cleans the \textit{dirty map} by applying deblurring graph convolutions and a 
 %at each step. 
The trainable parameters are $\theta$, $\mathbf{B}$ and $\tau$ (a bias term), and are shared across the $L$ iterations.
In practice, $F$ of these operations happen in parallel, one for each frequency band.
% (linearly from  1.5 to 4.5 kHz).
DeepWave's output acoustic image (or SIM) has $N$ pixels, proportional to the number of microphone channels.

\section{Approach}
\label{sec:methods}
\subsection{DoAE from acoustic images via K-means clustering}
\label{sec:DeepWavekmeans}

% \subsubsection{DeepWave+K-means}
% \label{sec:DeepWavekmeans}

First we optimized a DeepWave model to get a sense of maximum performance on the LOCATA dataset (see Methods). 
This model processes a high-resolution microphone.
We apply K-means clustering over DeepWave's SIMs to calculate the DoA.
% from sound intensities. 
For a 32ch input with $F=9$ and $L=5$, DeepWave generates a $N=242$ acoustic image with values ranging between 0 and 1. 
After arranging the $N$ pixels in a 2D space $N=A \times E$ for azimuth ($A$) and elevation ($E$) we obtain $\mathbf{I} \in \mathbb{R}^{F\times A\times E}_{[0, 1)}$. 
%, the square term is added to better highlight sound intensities.
Next, we apply a 1D Tukey window along the elevation axis with a $0.8$ tapering factor to remove artifacts closer to the poles. 
% Subsequently, we sum along the dimension of the $F$ frequency bands, yielding $\mathbf{I} \in \mathbb{R}^{A\times E}_{[0, 1)}$. 
Finally, we run K-means clustering with $K=3$ on the $15$ pixels with the maximal intensity (all others are clipped to zero).
This yields three centroids that represent DoA coordinates. 
% The centroid from each cluster provides the resulting DoAE coordinates.
To make this algorithm robust, we apply two post-processing heuristics: (1) ignore clusters where points are separated by more than $15^\circ$ from the centroid, and 
%away from their centroid (with a requirement of at least 2 points within this range) and 
(2) merge clusters with centroids within $15^\circ$ of each other.

\subsection{Upsampling $\mathbf{\Sigma}$ using super-resolution}    
% \section{A method for microphone array signal upsampling}
\label{sec:upsampling}

DeepWave's resolution is proportional to the number of microphone channels. 
%, denoted as $\mathbf{\Sigma}$. 
%Out initial attempts to employ DeepWave for DoAE with such low-resolution input $\mathbf{\Sigma}$ yielded suboptimal learning results. Therefore, w
Lower-resolution microphones, such as 4 channel, are commonly used in SELD datasets.
We introduce a channel upsampling method derived from the Deep Back Projection Network (DBPN) by Haris et al. \cite{haris2018deep}, a computer vision super-resolution model.
Our complex-valued DBPN (CDBPN) upsamples $\mathbf{\Sigma} \in \mathbb{C}^{4\times 4} \rightarrow \mathbf{\Sigma} \in \mathbb{C}^{32\times32}$ (i.e. from a 4ch to a 32ch array; a factor of 8). 

\subsection{DeepWave SIMs as inputs for DoAE}

\begin{figure}

\includegraphics[width=8.5cm]{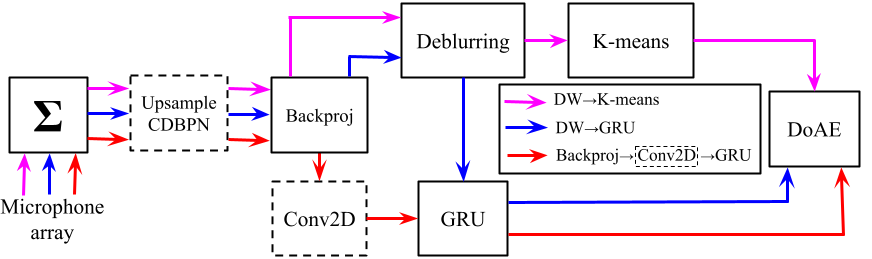}
\caption{Signal pipeline in the models we study.}
%the different variants of DeepWave for DoAE that we study.}
\label{fig:dw-diagram}
\end{figure}

The original DeepWave processes 100ms at a time and lacks temporal memory \cite{simeoni2019deepwave}. 
We also study the effect of adding a gated recurrent unit (GRU) on top.
In other words, DeepWave's SIM becomes the GRU input (Figure \ref{fig:dw-diagram}).
We train models in an end-to-end fashion using the ADPIT loss, which makes this type of model have the multi-ACCDOA representation to localize overlapping sound sources \cite{shimada2022multi}. 
% It is worth noting that, when en-to-end training all models and performing network dissection, we observe that the input to the GRU no longer is interpretable acoustic map as shown in Figure \ref{fig:dw}. Instead, the spherical acoustic image becomes a sound intensity embedding vector.

% Since $\mathbf{\Sigma}$ is complex-valued, our outputs model is a %we have devised a 
% complex-valued version of the . 
% This process estimates an equivalent $\mathbf{\Sigma} \in \mathbf{C}^{32 \times 32}$, as if it had been captured by a high-resolution array with 32 channels.

\section{Methodology}
\label{sec:methodology}

\subsection{Datasets}
\label{sec:dataset}
% \subsubsection{Simulated data}

% Sound events have instantaneous DoA labels (every 100ms).
\underline{Evaluations with real recordings.} We evaluate models on LOCATA because it was recorded with an EigenMike (32ch) in a room where human actors moved while speaking (the microphone also moves in some recordings) \cite{lollmann2018locata}.
To apply CDBPN upsampling we simulate a tetrahedral microphone (4ch) using the 6th, 10th, 22nd, and 26th EigenMike channels (following Adavane et al.'s \cite{adavanne2019localization} approach). 
Metadata does not indicate a speaker's gender, so annotations only have the instantaneous DoA (every 100ms).
Our study focuses on DoAE, so these annotations are all we need (i.e. we do not classify sound events into  categories). 
We also evaluate on STARSS23 \cite{shimada2023starss23}, which contains more rooms and uses a 4ch microphone.
We only use its DoA annotations, as we do not classify events into specific classes. % because we focus on DoAE. % sound event

%DeepWave'soriginally assumes an input captured with a high-resolution microphone array, e.g., EigenMike (32 channels) or Pyramic (48 channels). 
\underline{Simulated recordings for training.} 
To train models we simulate 32ch recordings.
%We modified the open-source data generation scripts by Politis and Krauze \cite{krauze2022generator}. 
Our simulated dataset was generated using the SpatialScaper library \cite{roman2024spatial}.
%They originally simulated tetrahedral (4ch) recordings using sparse room impulse responses (RIRs), which they convolve with sound waveforms to spatialize them.
We integrated two RIR databases that use the EigenMike. % of 32ch RIRs, which were collected with the % from two rooms two room impulse responses (RIRs) collected using an 
% EigenMike em32. 
The first one is %RIR data comes from the 
METU-SPARG, collected %using logarithmic sine sweeps measured 
at 240 points on a cubic grid %space of 0.5m sides and 0.3m vertical, 
% centered around an EigenMike 
in a classroom %dataset, collected by Olgun et al. 
\cite{orhun_olgun_2019_2635758}, and the second one is ARNI 
\cite{mckenzie2021dataset}, collected in the ``variable acoustics room'' at Aalto University. 
% %They placed three loud speakers around the center of a room, mimicking a stage and facing the south wall.
% The em32 RIRs for the first room come from the METU-SPARG dataset, 
% collected by Olgun et al. \cite{orhun_olgun_2019_2635758} using logarithmic sine sweeps measured at 240 points on a cubic grid %space of 0.5m sides and 0.3m vertical, 
% centered around an EigenMike in a classroom. %[What is the shape of those points?] %covering the whole azimuth range with approximately $\pm 50^\circ$ in elevation. <- this is not true. 
% The em32 RIRs for the second room come from the ARNI 
% dataset, collected by McKenzie et al. \cite{mckenzie2021dataset} in the Arni variable acoustics room at Aalto University. 
% %They placed three loud speakers around the center of a room, mimicking a stage and facing the south wall (see \cite{mckenzie2021dataset} for more details). In our study, we use the RIR data from five EigenMikes placed 1.3m from the south wall, each separated by 1.25m along the wall.
We simulate soundscapes with speech %We integrated these rooms in the open-source data generator to use their RIRs to convolve human speech samples 
from the FSD50K dataset \cite{fonseca2021fsd50k} spatialized in either of the two rooms. 
We generate a total of 50min of data in the METU room and 10min in ARNI. 

%The data generator produces azimuth and elevation labels for each sound event indicating every 100ms an event's location in azimuth and elevation. 

%Since we study upsampling from four to 32 channels, we generate
%Our development dataset consists of soundscapes generated using the RIRs form these two rooms convolved with 

Models evaluated on STARSS23 are trained using the companion dataset with simulated 4ch recordings using RIRs from 12 rooms \cite{politis_archontis_2022_6406873}. 
We refer to it as ``DCASE-sim''.%\footnote{released in 2022 for the Detection and Classification of Acoustic Scenes and Events (DCASE) SELD challenge}''.

%s have annotated azimuth and elevation every 100ms.

% [One sentence about the principle of DoAE data generation with IRs]
% [State that you use data from two rooms. First talk about the METU room. Who are the authors? What hardware do they use collect? What method do they use to collect the IRs? How many locations? Give a summary of how many IRs you use, what shape do they form?]
% [Next talk about the ARNI room. Who are the authors? What hardware do they use collect? What method do they use to collect the IRs? How many locations? Give a summary of how many IRs you use, what shape do they form?]
% [Talk about the SELD DCASE data generator. How it works. What you did to make it work with METU+ARNI].
% [How many tracks did you generate for each. How many hours did you have in the end? Did the sources overlap?]

%with a total of 13 target sound event classes. We use the "development" set, which comes with includes DoAE labels the both train and test splits.

\begin{table*}
\begin{center}
\begin{tabular}{llllll}
\toprule
Input & \textbf{Model / operations} & $LE$ $\downarrow$ (std.) & $LR$ $\uparrow$ (std.)\\ 
\midrule\midrule

    \multirow{4}{*}{32ch} & Backproj $\rightarrow$ Deblurring $\rightarrow$ K-means & \bf{14.8}$^o$ ($\pm 0.00$)& \bf{99.20} ($\pm 0.00$)\\ \cline{2-4}

    % Up+DW+K & \multicolumn{1}{l}{tetra} & 27.10  &  \textbf{99.2} \\ \hline
    
    & Backproj $\rightarrow$ Deblurring $\rightarrow$ GRU & 20.9$^o$ ($\pm 3.63$) & 69.36 ($\pm 2.15$) \\ \cline{2-4}

    & Backproj $\rightarrow$ GRU & 20.0$^o$ ($\pm 3.01$) & 70.46 ($\pm 4.14$) \\ \cline{2-4}

    & Backproj $\rightarrow$ Conv2D $\rightarrow$ GRU & 15.96$^o$ ($\pm 2.51$) & 76.36 ($\pm 9.21$) \\ \hline\hline

    \multirow{4}{*}{4ch $\rightarrow$ CDBPN $\rightarrow$ 32ch} & Backproj $\rightarrow$ Deblurring $\rightarrow$ K-means & 27.10$^o$ ($\pm 0.00$) & \bf{99.20} ($\pm 0.00$)\\ \cline{2-4}

    % Corr+G & em32 & 180 & 0 \\ \hline

    & Backproj $\rightarrow$ Deblurring $\rightarrow$ GRU & 20.30$^o$ ($\pm 2.76$) & 72.33 ($\pm$ 0.04) \\ \cline{2-4}

    & Backproj $\rightarrow$ GRU & 18.06$^o$ ($\pm 3.53$) & 70.96 ($\pm 3.82$) \\ \cline{2-4}

    & Backproj $\rightarrow$ Conv2D $\rightarrow$ GRU & \bf{17.42}$^o$ ($\pm 0.73$)  & 82.40 ($\pm 5.76$)
    
    % Up+Backproj+G & \multicolumn{1}{l}{tetra} & \underline{16.59} ($\pm 1.47$) & 76.03 ($\pm 1.4$) \\ \hline
    
    % Up+Corr+G & tetra & 180 & 0 

    \\\hline\hline
    % SELDnet2322    & tetra & 25.83 ($\pm 2.27$) & 85.66 ($\pm 6.44$) \\ 
    % \hline
    % SELDnet2322    & FOA & 20.60 ($\pm 3.52$) & 70.23 ($\pm 0.03$)  \\ 
    % \hline
    4ch  & SELDnet23 & 16.8$^o$ ($\pm 0.45$) & 77.13 ($\pm 3.84$) \\ 
    \hline
    FOA & SELDnet23 & 22.43$^o$ ($\pm 3.80$) & 80.73 ($\pm 10.98$)  \\ 
    \hline
    % Hu-IACAS\cite{} & FOA+tetra & 19.83 ($\pm 0.75$) & 80.66 ($\pm 7.26$)     \\ \hline
    FOA + 4ch & EINV2 & 19.83$^o$ ($\pm 0.75$) & 80.66 ($\pm 7.26$) \\
    % & & \multicolumn{1}{r}{} &  \\
   \bottomrule
        \end{tabular}
    \caption{Localization error (LE) and recall (LR) on LOCATA. Scores reflect the average across three experimental replications. Note that the EINV2 model uses both tetrahedral microphone (4ch) and first order ambisonics (FOA) inputs.}% (standard deviation in parenthesis).} %Double lines separate 1) using all EigenMike 32ch, 2) using 4ch upsampled to 32ch, 3) baseline (SELDnet2323) and SoTA (EINV2) models.}
    \label{tab:loc_results}
    
\end{center}
\end{table*}

\subsection{Models, ablations, and baselines}
% \subsection{Ablation study}

% First, we benchmark our proposed models LOCATA dataset (32ch) to identify the optimal one. 
The two main models we study are DeepWave+K-means and DeepWave+GRU. 
We carry out an ablation study of the DeepWave+GRU model\footnote{the original DeepWave paper has an ablation study without the GRU \cite{simeoni2019deepwave}} to understand the role of the deblurring operation given the additional GRU. %and whether it is unnecessary given the added GRU.
% This will allow us to identify potential redundancies between operations carried out by the combined DeepWave and GRU.
Therefore we have three variants: 1) Backproj $\rightarrow$ Deblur $\rightarrow$ GRU, 2) Backproj $\rightarrow$ GRU, and 3) Backproj $\rightarrow$ Conv2D $\rightarrow$ GRU (i.e. we replace deblurring with a 2D convolution layer)\footnote{The variant that passes $\mathbf{\Sigma}$ directly to the GRU does not learn, independent of whether CDBPN upsampling is used.}.
% We also evaluate whether a convolution layer after the backprojection operation 
% The variant without the GRU is equivalent to the original DeepWave.
% Ablating the DeepWave+K-means model is redundant with the ablation study carried out in , where they found all operations in DeepWave to be necessary for optimal acoustic imaging.

Our optimizations are gradual steps to enable DoAE using a DeepWave backbone. DeepWave+K-means enables DoAE on SIMs, provided high-resolution array data. CDBPN introduces low-resolution array upsampling. A GRU on top of a DeepWave backbone allows end-to-end training using multi-ACCDOA representations. In each variant DeepWave operations are key for DoAE.

Across datasets we compare performance against two models: SELDnet23, a baseline for DoAE, and EINV2, the highest-ranked open-source model\footnote{acording to the ``DCASE'' 2023 SELD challenge}. 
Note these models were designed to carry out SELD (i.e. classification in addition to DoAE). 
Also, EINV2 assumes data augmentation, but the authors did not make this code available on their repository\footnote{https://github.com/Jinbo-Hu/DCASE2022-TASK3}.
Therefore we use the EINV2 without data augmentation. 

% At the end, we show that DeepWave can perform the DoAE task on two different and commonly-used SELD datasets featuring low-resolution tetrahedral microphone recordings (i.e. our results on STARSS23 and LOCATA datasets). Our study includes a systematic benchmark comparing DeepWave’s performance against the SELDnet and EINV2 models. Therefore, our experimental methodology includes an evaluation of DeepWave’s generalization as a model for DoAE tasks.

\subsection{Training procedure and evaluation metrics}

\underline{CDBPN:} We train CDPBN to upsample from 4ch to 32ch. 
We use METU soundscapes for training and ARNI for validation.
The input and target are the instantaneous (100ms) 4ch $\mathbf{\Sigma}\in\mathbb{C}^{4\times 4 \times F}$ and 32ch $\mathbf{\Sigma}\in\mathbb{C}^{32\times 32 \times F}$, respectively.
% The target output is derived from the corresponding 32-channel track.
We use the original DBPN code, but we use $F$ channels instead of RGB, and train two models, one for the real part and another for the imaginary part of $\mathbf{\Sigma}$.
We freeze the optimal CDBPN and use it in experiments where we upsample from 4ch to 32ch.  

\underline{LOCATA experiments:} We train models using the generated METU soundscapes, and we validate using the ARNI ones. 
The final evaluation is carried out across the entire LOCATA dataset. 
Frequency bands are nine total ($F=9$) and linearly spaced from 1.5kHz to 4.5kHz.
%, consistent with the original DeepWave implementation.
We evaluate DeepWave models with and without the CDBPN upsampling.

\underline{STARSS23 experiments:} 
STARSS23 comes divided in four directories (``dev-train-tau'', ``dev-train-sony'', ``dev-test-tau'' and ``dev-test-sony''),
% The ``tau'' files were recorded in Finland, while the ``sony'' ones in Japan. 
each with a unique set of rooms, and each room has its own sound sources (i.e. a set of ``actors'' and sounding objects such as a guitar or household blender). 
We train models using ``DCASE-sim'', ``dev-train-tau'', and ``dev-train-sony'', and we validate using ``dev-test-tau''.
The final evaluation is carried out on the ``unseen'' rooms in ``dev-test-sony''.
% Frequency bands are sixteen total ($F=16$) and logarithmically spaced from 50Hz to 4.5kHz, consistent with the SELDnet23 and EINV2 frequency bands. 
Note that the optimal DeepWave model that we evaluate on the STARSS23 dataset uses CDBPN since the STARSS23 and ``DCASE-sim'' datasets only have 4ch. 

\underline{Metrics.} We evaluate
%To evaluate model performance we use
% the DCASE challenge Task 3 localization metrics \cite{politis2020overview}: 
Localization Error (LE), the radial difference between predicted and true location per sound event, and Localization Recall (LR), the true positive rate of instantaneous detections out of the total annotated sound event instances \cite{politis2020overview}. 
We do not measure sound classification performance since we focus on sound localization. % based on category of sounds Since we do not clarssify sounds, w

% The tetra4ch dataset assumes channels $6, 10, 26, 22$ (1 index) from the em32ch dataset.

% using the spatial soundscapes we generated using the METU RIRs, using the ARNI RIRand validate this training with the ARNI RIR
% All models and experiments use the same training, validation, and test sets in this study were trained using equivalent development and evaluation datasets in both em32ch and tetra4ch format. 

% We leverage both datasets to train the CDBPN upsampling model, assuming the tretra4ch data as the low-resolution features and the em23ch data as the super-resolution targets.

% For Table \ref{tab:loc_results}, the model training utilized the METU room data for training, ARNI room data for validation, and LOCATA dataset for evaluation. 

% In Table \ref{tab:starss_results}, the results were obtained by training on STARSS23 , with validation on STARSS23 "dev-test-tau" and evaluation on "starss-dev-sony." This modified the DeepWave hyperparameters, incorporating 16 logarithmically spaced frequency bands ranging from 50Hz to 4.5kHz to capture the most frequency ranges of classes in STARSS23.

\section{Results}
\label{sec:results}

\begin{figure}

\includegraphics[width=8.4cm]{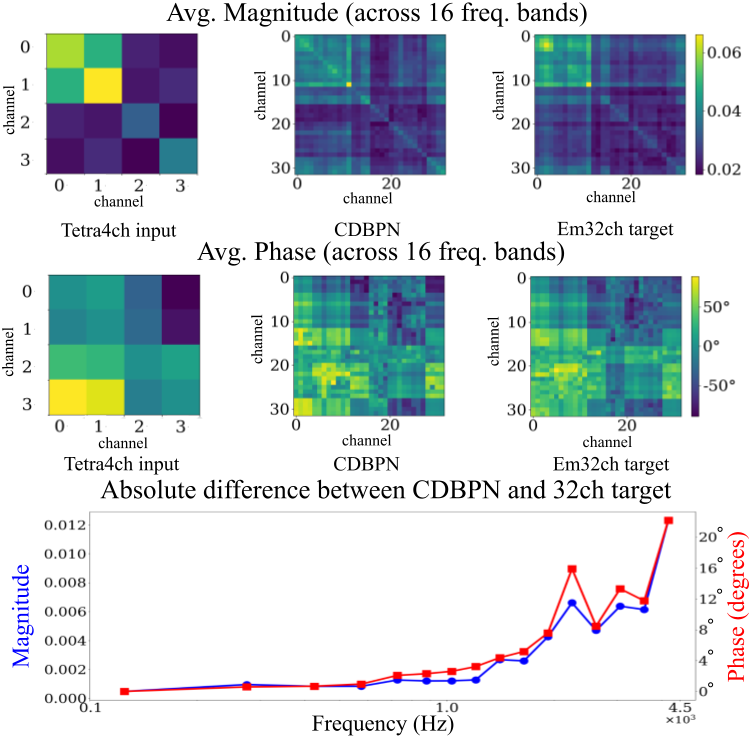}
\caption{CDBPN upsampling of a single reverberant white noise source directly facing the front of the microphone.}
\label{fig:cdbpn}
\end{figure}

Table \ref{tab:loc_results} shows performance by the model variants on the LOCATA dataset and compares against SELDnet23 and EINV2. DeepWave+K-means with 32ch a input outperforms all models with low LE and high LR. 
We optimized this model to give a sense of ``ceiling'' performance.
Its counterpart using 4ch upsampled with CDBPN shows a deteriorated LE (see Figure \ref{fig:cdbpn} bottom), but LR is same. 
We enhanced CDBPN for upsampling $\mathbf{\Sigma}$; however, this does not encompass SIM generation, thus posing challenges for K-means post-processing.

%However, the high LR indicates that activation to sound events remains equally responsive.
%%%
%%% TODO: SHOW IMAGE
%%%
% spreads the clusters centroi acoustic image loci, , but activation as a function of sound  to detect sounds remains unaffected. 
% This could be due to the fact that DeepWave's output acoustic image is deteriorated when CDBPN upsampling is used. %a low-pass version of the original one when using the upsampling method, thus becoming harder to cluster relevant intensities via K-means. % IRAN: should we include an image showing this qualitatively?

The models with Backprojection, Deblurring, and GRU 
%DeepWave+GRU model models that use DeepWave back-projection followed by 5 deblurring graph convolutions and a GRU 
show LE around 20$^o$, independent of whether the input is 32ch or upsampled from 4ch using CDBPN. 
This indicates that the GRU is resilient to CDBPN distortions, and can improve LE by \textasciitilde7$^o$ compared to DeepWave+K-means.
% For the specific case of input that is upsampled from 4ch to 32ch, results indicate that the GRU
% serves as a
% These results shows that a GRU can take DeepWave acoustic image as input to carry out DoAE with an ACCDOA representation. 
%backbone to carry out the DoAE regression task. 

The models that only use Backprojection and GRU further improve LE, indicating that Deblurring is unnecessary given the GRU. 
This is consistent with the result we discussed in the previous paragraph, where we note that the GRU is resilient to CDBPN distortions. 
In other words, DeepWave's Deblurring becomes unnecessary when the GRU is added. 
Adding a Conv2D layer between Backprojection and the GRU further improves performance.
This makes sense, as the Conv2D layer operates across the $F$ frequency ``channels'', and passes the GRU a representation with encoded information across frequency bands. 
In other words, it saves the GRU the step of having to aggregate information across the frequency axis.

Note that all GRU models show a deteriorated LR compared to the ones using K-means. 
This could be caused by the temporal memory that the GRU adds, which may come with a time constant to ``react'' in response to sounds appearing and disappearing from the scene. 
In contrast, DeepWave+K-means operated on single frames of audio without memory, and was optimized to ``quickly'' react to sound activity.

\begin{table}
 \begin{center}
 \begin{tabular}{llcc}
   \toprule\toprule
    Input & \textbf{Model} &  $LE$ $\downarrow$ &  $LR$ $\uparrow$ \\ 
   \midrule\midrule
    % &    Mean  & Median & Std. deviation \\ \hline
    % tetra4ch $\rightarrow$ up32ch & Backproj $\rightarrow$ GRU& \textbf{23.0$^o$} &  72.5 \\ \hline
    up32ch & Backproj $\rightarrow$ Conv $\rightarrow$ GRU$^*$ & \textbf{20.5$^o$} &  70.1 \\ \hline
    4ch & SELDnet23 & 23.3$^o$  & 82.3    \\ \hline
    FOA & SELDnet23 & 21.9$^o$  & 83.0    \\ \hline
    FOA + 4ch & EINV2 & 24.0$^o$  & \textbf{84.2}    \\ \hline
   \bottomrule
   \end{tabular}
\end{center}
\setlength{\belowcaptionskip}{-15pt}
 \caption{Performance of SELDnet23, EINV2 and our best model on the STARSS23 ``dev-test-sony'' split. $^*$Plus two MHSA layers after the GRU.}
 \label{tab:starss_results}
\end{table}

Next we evaluated the optimal model (CDBPN $\rightarrow$ Backproj $\rightarrow$ Conv2D $\rightarrow$ GRU) on STARSS23. 
Compared to LOCATA, STARSS23 contains more diverse sound sources and acoustic conditions. We found optimal performance by adding two multi-headed self-attention (MHSA) layers after the GRU \cite{sudarsanam2021assessment}. Our model outperforms the LE performance of SELDnet trained with FOA, as well as SELDnet trained with 4ch and EINV2. %(EINV2 shows the best LR).
% Interestingly, SELDnet23 and EINV2 show comparable performance. 
Reasons for the deteriorated EINV2 performance compared to SELDnet23 include the smaller validation set we used (``dev-test-tau''), and the test set being smaller and more challenging (``dev-test-sony'').
Note that EINV2 is known to have a female speech LE ``a lot higher than average'' \cite{hu2022}.
%The EINV2 authors in their paper mention that EINV2 has a female speech LE ``a lot higher than average'' \cite{hu2022}. %, which may explain the performance degradation compared to SELDnet23 and other DeepWave-based models.
% Table \ref{tab:starss_results} compares the performance of the baseline SELDnet23 model compared to the  Upsampled Back-projection+GRU network (Up+Backproj+G) using the STARSS2023 dataset \cite{politis_adavanne_2022_6387880}. 

When it comes to CDBPN, we analyzed its performance in terms of the magnitude and phase error.
Figure \ref{fig:cdbpn} breaks down these results. 
The top matrices show the input $\mathbf{\Sigma}_4\in\mathbb{C}^{4\times 4 \times F}$, CDBPN output $\mathbf{\hat{\Sigma}}_{32}\in\mathbb{C}^{32\times 32 \times F}$ and target $\mathbf{\Sigma}_{32}\in\mathbb{C}^{32\times 32 \times F}$ (averaged across the frequency axis) for magnitude (first row) and phase (second row).
The bottom plot shows the magnitude and phase error between $\text{avg}(\mathbf{\hat{\Sigma}}_{32})$ and $\text{avg}(\mathbf{\Sigma}_{32})$ as a function of frequency (16 logaritmically-spaced bands). 
Results indicate that upsampling $\mathbf{\Sigma}$ at higher-frequency bands is more challenging for CDBPN.
% The signal used to generate these plots was reverberant white noise from a single source directly facing the microphone.
% Finally, the bottom panel shows DeepWave's acoustic image when using CDBPN upsampling to processing the same audio as in Figure \ref{fig:dw}.

% \vspace{-0.5cm}
\section{Conclusion and future work}
\label{sec:conclusion}

We have studied SIMs for DoAE.
We focused on DeepWave because of its high-resolution SIM.
We proposed a CDBPN model to upsample microphone array information from 4ch to 32ch. 
This allows our optimal model (4ch $\rightarrow$ CDBPN $\rightarrow$ 32ch $\rightarrow$ Backproj $\rightarrow$ Conv2D $\rightarrow$ GRU) to  process existing SELD data with 4ch audio. 
% We outperformed SELDnet23 and EINV2, the top-performing open-source DoAE model.
The model can be interpreted as the combination of DeepWave (Backprojection to generate a SIM) and SELDnet23 (Conv2D + GRU) operations. Our systematic benchmark against the SELDnet and EINV2 models on the LOCATA and STARSS dataset is an indicator that our models can be generalized to other DoAE tasks.

Future work includes developing our model to also carry out sound event classification.
Furthermore, generalizing CDBPN work with arbitrary microphone array shapes will be an important endeavor. 
For the time being, our results demonstrate the advantages of using SIMs for DoAE, which could benefit existing and future DoAE and SELD models. 

\section{Acknowledgements}
This work is supported by the National Science Foundation grant no. IIS-1955357. The authors thank the funding source and their grant collaborators.

\vfill\pagebreak

\bibliographystyle{IEEEbib}
\let\oldbibitem\bibitem
\def\bibitem{\vspace{-1.0ex}\oldbibitem}
\bibliography{refs, strings}
\end{document}